\documentstyle[preprint,prl,aps]{revtex}
\begin{document}
\draft
\title{CLUSTER INTERPRETATION OF  PROPERTIES OF  ALTERNATING
PARITY BANDS IN HEAVY NUCLEI}
\author{T.M.Shneidman$^{a,b}$, G.G.Adamian$^{a,b,c}$, N.V.Antonenko$^{a,b}$,
R.V.Jolos$^{a,b}$ and W.Scheid$^{a}$}
\address{$^{a}$Institut f\"ur Theoretische Physik der
Justus--Liebig--Universit\"at,
D--35392 Giessen, Germany\\
$^{b}$Joint Institute for Nuclear Research, 141980 Dubna, Russia\\
$^{c}$Institute of Nuclear Physics,
Tashkent 702132, Uzbekistan}
\date{\today}
\maketitle
\begin{abstract}
The properties of the states of the alternating parity bands in actinides,
Ba, Ce and Nd isotopes are analyzed
within a cluster model. The model is based on the assumption that cluster
type shapes are produced by the collective motion of the nuclear system in the mass
asymmetry coordinate. The calculated
spin dependences of the parity splitting and of the
electric multipole transition moments are  in  agreement
with the experimental data.
\end{abstract}

\section{Introduction}
The low-lying negative parity states observed in actinides and in heaviest known
Ba, Ce, Nd and Sm isotopes are definitely related to reflection--asymmetric shapes
\cite{ahmad,butler}. There are several approaches to treat collective motion
leading to reflection--asymmetric deformations.
One of them is based on the concept of a nuclear mean
field which has a static
octupole deformation or is characterized by  large amplitudes
of reflection--asymmetric
vibrations around the equilibrium shape
\cite{butler,Moeller,Leander,Sobic,Robledo}.
In this approach the parity splitting is explained by octupole deformation.
Another approach \cite{Iachello,Gai,Daley,Daley1} is based on the
assumption that the reflection--asymmetric shape is a consequence of
alpha--clustering in nuclei
\cite{Wilder,Bromley,Cindro}. In the algebraic model
\cite{Iachello,Gai,Daley,Daley1} the corresponding wave functions of the ground and
excited states consist of components without and with dipole bosons
(in addition to the quadrupole bosons), which are related to mononucleus and
alpha--cluster components, respectively.
The variant of algebraic model including the octupole bosons in
addition to the dipole bosons has been applied in \cite{Zamfir,Zamfir1}
to the description of the low-lying negative parity states in actinides.
In \cite{Buck,Buck1,Buck2,Buck3}
a cluster configuration with a lighter cluster heavier than $^{4}$He
was used in order to describe the properties of the low--lying positive
and negative parity states. In both models
\cite{Iachello,Gai,Daley,Daley1}
and \cite{Buck,Buck1,Buck2,Buck3} the relative distance between the centers of mass of
clusters at fixed mass asymmetry is the main collective coordinate for the
description of the alternating parity bands.

Nuclear cluster effects are mostly pronounced in the light even--even $N=Z$ nuclei
with alpha--particle as the natural building block. There is a nice relationship
between alpha--cluster description and deformed shell model \cite{Wilder}.
It is known from Nilsson--Strutinsky type calculations for light nuclei that
nuclear configurations corresponding to the minima of the potential energy
contain particular symmetries which are related to certain cluster structures
\cite{Rae,Freer1,Freer2}. By using antisymmetrized molecular dynamics approach
\cite{Horiuchi1,Horiuchi}, the formation and dissolution of clusters in light nuclei, like
$^{20}$Ne and $^{24}$Mg, are described. The idea of clusterization applied to heavy
nuclei does not contradict the mean field approach. The coexistence of the
clustering and of the mean field aspects is a unique feature of nuclear many body
system. The problem of existence of a cluster structure in a ground state of heavy
nuclei has attracted much attention, especially, because of the experimentally observed
cluster decay \cite{Zam}. The available experimental and theoretical results
provide the evidence for existence of fission modes created by the clustering of the
fissioning nuclei \cite{Pashkevich}.
Indications of
clusterization of highly deformed nuclei are demonstrated in \cite{Aberg,sdhd}.

The aim of the present paper is a development of the cluster--type model which provides not only a
qualitative but also a quantitative explanation of the properties of
alternating parity bands. The description of the excitation spectra, $E\lambda$--transition
probabilities ($\lambda$=1,2,3) and the angular momentum dependence of the
parity splitting \cite{Jolos1,Jolos2} are the main subjects of this paper. Our
model is based on the assumption that the reflection--asymmetric shapes
are produced by the collective motion of the nuclear system in the mass asymmetry
coordinate \cite{Shneidman}. The values of the odd multipolarity transitional
moments (dipole and octupole) are  strongly correlated with the mass asymmetry
deformation of nucleus. In general, the value of the quadrupole moment is related to the degree
of the quadrupole correlations (deformation) in nucleus. However, the collective
motion in the mass asymmetry degree of freedom simultaneously creates a deformation
with even and odd multipolarities. Therefore, the calculations of $E\lambda$--transition
moments are of interest in the proposed model. The single particle degrees of
freedom are not
taken explicitly into consideration since our aim is to show that the suggested cluster model
gives a good quantitative explanation of the observed properties
of the low--lying negative parity states. If it is so, this model can serve as a
good ground for development of an extended model with additional degrees of freedom.

It should be noted that the first results of calculations
of the alternating parity
spectra for a few actinides within the cluster
model are already presented in the Letter \cite{Shneidman}.
Besides Ra, Th and U isotopes, in the present paper
we present the results of
calculations of the energies of alternating parity bands
in $^{240,242}$Pu,
       $^{144,146,148}$Ba, $^{146,148}$Ce, and $^{146,148}$Nd.
The electromagnetic transitions are described in this paper with the
cluster model for many nuclei and
the spin dependence of the intrinsic
        quadrupole transition moment is predicted for $^{238}$U.
Simple analytical
expressions obtained for the parity splitting and
the spectra of alternating parity bands
are useful for the estimations.
The dependence of alpha--clusterization in actinides on
the angular momentum is shown for the first time.

\section{Model}

\subsection{Hamiltonian in mass asymmetry coordinate}

Dinuclear  systems consisting of a heavy cluster  $A_1$ and
a light cluster $A_2$
were first introduced to explain data on
deep inelastic and fusion reactions with heavy ions
\cite{volkov,obzor,isot}.
The mass asymmetry coordinate $\eta$, defined as
$\eta=(A_1-A_2)/(A_1+A_2)$,
($|\eta|$=1 if $A_2$=0 and $A_1=A$), which describes a partition of nucleons
between the nuclei forming the dinuclear system
and the distance $R$ between
the centers of clusters
are used as  relevant collective variables \cite{Greiner}.
The wave function in $\eta$
can be thought as a superposition of different
cluster--type configurations
including the mononucleus configuration with $|\eta|$=1, which are realized
with certain probabilities.
The relative contribution of each cluster component
in the total wave function is determined by
the collective Hamiltonian described below.
Our calculations have shown that
in the considered cases the dinuclear configuration
with an alpha cluster ($\eta=\eta_{\alpha}$)
has a potential energy which is close or even smaller
than the energy of the mononucleus at $|\eta|=1$
\cite{sdhd,Shneidman}.
Since the energies of configurations with a light cluster heavier
than an $\alpha$--particle
increase rapidly with decreasing $|\eta|$,
we  restrict our investigations to
configurations with light clusters not heavier than Li
($\eta=\eta_{Li}$), i.e. to
cluster configurations near $|\eta|=1$.

The Hamiltonian describing the dynamics in $\eta$ has the following form
\begin{eqnarray}
\label{ham1}
H=-\frac{\hbar^2}{2 B_\eta}\frac{d^2}{d\eta^2}+U(\eta ,I),
\label{ham_eq}
\end{eqnarray}
where $B_{\eta}$ is the effective mass and $U(\eta ,I)$ is the potential.
In order to calculate the dependence of parity splitting
on the angular momentum and the electric dipole, quadrupole and octupole
transition moments we search for solutions of the stationary Schr\"odinger
equation describing the dynamics in $\eta$:
\begin{eqnarray}
\label{ham2}
H\Psi_n(\eta,I)=E_n(I)\Psi_n(\eta,I).
\label{schr_eq}
\end{eqnarray}
The eigenfunctions $\Psi_n$ of  this Hamiltonian have a well
defined parity with respect to the reflection $\eta \to -\eta$.
Before we come to the results of Eq.~(\ref{schr_eq}),
we discuss the calculation of the potential $U(\eta,I)$,
the mass parameter $B_\eta$ and the moments of inertia
$\Im (\eta )$ appearing in $H$.

\subsection{Potential energy}

The potential $U(\eta,I)$ in Eq.~(\ref{ham_eq})
is taken as a dinuclear potential energy for $|\eta |< 1$
\begin{eqnarray}
U(R, \eta, I)= B_1(\eta)+ B_2(\eta)-B + V(R=R_m,\eta,I).
\label{pot_eq}
\end{eqnarray}
Here, the internuclear distance $R=R_m$ is the touching distance
between the clusters and is set
to be equal to the value corresponding
to the minimum of the potential in $R$ for a given $\eta$.
The quantities $B_1$ and  $B_2$ (which are negative) are the experimental binding
energies of the clusters forming the dinuclear system  at a given mass
asymmetry $\eta$, and $B$ is the binding energy of the mononucleus.
The quantity $V(R,\eta,I)$ in (\ref{pot_eq}) is the
nucleus-nucleus interaction potential.
It is given as
\begin{eqnarray}
\label{potential}
V(R,\eta,I)=V_{coul}(R,\eta) + V_N(R,\eta) + V_{rot}(R,\eta,I)
\label{V_eq}
\end{eqnarray}
with the Coulomb potential $V_{coul}$, the centrifugal potential
$V_{rot}=\hbar^2 I(I+1)/(2\Im (\eta,R))$
and the nuclear interaction $V_N$.
In the realization of the cluster model developed in this paper, where
overlap of clusters is much smaller than in the model of \cite{Buck},
the choice of the relevant cluster configuration follows the minimum
of the total potential energy of the system with a cluster--cluster interaction taken
additionally into consideration. As the result we describe the same nuclear
properties as in \cite{Buck} with configurations of clusters having larger mass
asymmetry and a smaller overlap.

The potential $V(R,\eta,I)$ and the moment of inertia
$\Im (\eta,R)$) are calculated for special cluster configurations only,
namely for the mononucleus ($|\eta|$=1) and for the
two cluster configurations with the
$\alpha$-- and Li - clusters as light clusters, respectively.
These calculated points are used later to interpolate the potential
smoothly  by a polynomial.
The energies of the Li-cluster configurations are about 15 MeV larger than
the binding energies of the mononuclei considered. Therefore, for small
excitations only oscillations in $\eta$ are of interest
which lie in the vicinity
of $|\eta|$=1, i.e. only  cluster
configurations up to Li - clusters need to be considered.
The  potential
$V_N$ is obtained with a double folding procedure
with the ground state
nuclear densities of the clusters.
Antisymmetrization between the nucleons belonging to
different clusters is
regarded by a density dependence
of the nucleon--nucleon force which gives a repulsive core in
the cluster--cluster interaction potential.
Details of the calculation of $V_N$ are given in \cite{poten}.
The parameters of the nucleon-nucleon interaction are
fixed in nuclear structure calculations \cite{Migdal}.
Other details are presented in \cite{Shneidman}.

Our calculations show that the potential
energy has a minimum at $|\eta |$=$\eta_{\alpha}$ in
$^{218,220,222,224,226}$Ra and  $^{222,224,226}$Th
isotopes. In order to
demonstrate the dependence of the potential on the neutron number,
we present in Fig.~1
calculated values of $U(\eta_\alpha,I=0)\equiv U(\eta_\alpha)$
of configurations with an $\alpha$ - cluster
taking the long chain of  Ba isotopes as an example.
In the neutron deficient
isotopes $U(\eta_\alpha)$ is smaller than zero
and an $\alpha$-clusterization is more likely. When the neutron
number approaches the magic value of 82, the nucleus becomes
stiffer with respect to  vibrations in $\eta$ and $U(\eta_\alpha)$
is larger than zero. The appearance of  two
neutrons above shell closure is in  favor for an $\alpha$ - clusterization.
In this case $U(\eta_\alpha)$ drops much and again becomes smaller than zero.
Further addition of neutrons increases the nuclear
stiffness with respect to $\eta$  vibrations.

\subsection{Moments of inertia}

The calculation of the moment of inertia $\Im(\eta)=\Im (\eta,R_m)$ needed to
determine
the potential energy at $I\ne 0$ has been described in \cite{Shneidman}. For
completeness, we repeat in this subsection the most important information.
As
was shown in \cite{sdhd}, the highly deformed states are well described as
cluster systems and their moments of inertia  are about 85$\%$ of the
rigid-body limit.
Following this, we assume that the moment
of inertia of the cluster configurations with $\alpha$ and Li as light
clusters can be  expressed as
\begin{eqnarray}
\Im(\eta)=c_1(\Im^r_1+\Im^r_2+m_0\frac{A_1 A_2}{A}R_m^2).
\end{eqnarray}
Here, $\Im^r_i, (i=1,2)$ are the rigid body moments of
inertia for the clusters of the dinuclear system,
$c_1$=0.85 \cite{sdhd,Shneidman}
for all considered nuclei and  $m_0$ is the nucleon mass.

It should be noted  that
the angular momentum is treated in this paper
as the sum of the angular momentum of the collective rotation
of the heavy cluster and of the orbital momentum of the relative motion of the
two clusters.
Single particle effects, like alignment of the
single particle angular momentum in the heavy cluster, are presently
disregarded.

For $|\eta|=1$, the value of the moment of inertia
is not known from the data because
the experimental moment of inertia is a mean
value between the moment of inertia of the mononucleus ($|\eta|$=1)
and the ones of the cluster configurations arising
due to the oscillations in $\eta$.
We assume that
\begin{eqnarray}
\Im(|\eta|=1)=c_2\Im^r(|\eta|=1),
\end{eqnarray}
where $\Im^r$ is the rigid body moment of inertia of the mononucleus
with $A$ nucleons calculated with deformation parameters from
\cite{mollernix} and $c_2$ is a
scaling parameter which is fixed by
the energy of the first 2$^+$ or
other positive parity state, for example 6$^+$.
The chosen values of  $c_2$  vary in the interval $0.1<c_2<0.3$.
So, in our calculations there is a free parameter  $c_2$. However, this
parameter is used to describe
the rotational energies averaged over the parity and not the parity
splitting studied in this paper.

\subsection{Mass parameter}

The method of the calculation of the inertia coefficient $B_\eta$ used in this
paper is given in \cite{mass}. Our calculations show that $B_\eta$ is a smooth
function of the mass number $A$.
As a consequence, we take nearly the same
value of $B_\eta$=20 $\times10^4 m_0$ fm$^2$ for almost
all considered actinide nuclei with a variation of 10$\%$.
However, for
$^{222}$Th and $^{220,222}$Ra we varied $B_\eta$ in the range
$B_\eta$=(10-20)$\times10^4m_0$ fm$^2$ to obtain the correct
value of $E_0(I=0)$.
These variations of $B_\eta$ lead to better results
for light Ra isotopes than those in \cite{Shneidman},
where the obtained values of the parity
splitting at the beginning of the alternating parity band are smaller than
the experimental ones. Using a smooth mass dependence of $B_{\eta}$ \cite{mass}
we get $B_{\eta}$=4.5$\times 10^4 m_0$ fm$^2$ in the Ba, Ce and Nd region.
However, better results we obtain for $B_{\eta}$=3$\times 10^4 m_0$ fm$^2$.

For very asymmetric dinuclear systems, we can use simple analytical
expressions to establish a connection between the relative distance
and mass asymmetry coordinates on one side and the multipole expansion
coefficients $\beta_2$ and $\beta_3$ on the other side \cite{sdhd}
\begin{eqnarray}
\beta_2=\sqrt{\frac{5}{4 \pi}}\frac{\pi}{3}(1-\eta^2)\frac{R^2}{R_0^2},
\nonumber\\
\beta_3=\sqrt{\frac{7}{4 \pi}} \frac{\pi}{3}\eta (1-\eta^2)\frac{R^3}{R_0^3}.
\end{eqnarray}
Here, $R_0$ is the spherical equivalent radius of the corresponding
compound nucleus. One finds
\begin{eqnarray}
\frac{d\beta_3}{d\eta}=\frac{\sqrt{7\pi}}{12}
\left [(1+\eta)^{1/3}+(1-\eta)^{1/3} \right]^3 \nonumber \\
\times\left [
(1-3 \eta^2)+\eta(1-\eta^2)\frac{(1+\eta)^{-2/3}-(1-\eta)^{-2/3}}
{(1+\eta)^{1/3}+(1-\eta)^{1/3}}
\right].
\end{eqnarray}
In the actinide region for an $\alpha$ -particle configuration, $\eta \approx$0.96
and $(d\beta_3/d\eta )^2 \approx$11.25. Then the mass parameters for $\beta_3$
and $\eta$-variables are related as
\begin{eqnarray}
B_\eta \approx (d\beta_3/d\eta )^2 B_{\beta_3}.
\end{eqnarray}
If we take
the value of
 $B_{\beta_3}=200 \hbar^2$ MeV$^{-1}$
known from the literature  \cite{Leander}, then
$B_\eta \approx$9.3 $\times$10$^4 m_0$ fm$^2$.
This value is compatible with the one used in our calculations.

\section{Intrinsic electric multipole moments}

Solving the eigenvalue equation (\ref{schr_eq}),
we obtain the wave
functions of the positive and negative parity states for different values of
the quantum number $I$ of angular momentum. These wave functions
are used then to calculate
transition matrix elements of the electric multipole operators by
integration over $\eta$.
The electric multipole operators for a system of a dinuclear
shape have been calculated \cite{sdhd} by using the following expression
\begin{eqnarray}
\label{multipole1}
Q_{\lambda\mu}=
\sqrt{\frac{16\pi}{2\lambda +1}}
\int\rho^Z({\bf r})r^{\lambda}Y_{\lambda\mu}(\Omega)d\tau.
\end{eqnarray}
For slightly overlapping clusters when the intercluster distance $R_m$ is about
or larger
than the sum of the radii of clusters $(R_1 +R_2)$, the nuclear
charge density
$\rho^Z$ can be taken as a sum of the cluster charge densities
\begin{eqnarray}
\label{multipole2}
\rho^Z ({\bf r})=\rho_1^Z ({\bf r})+
\rho_2^Z ({\bf r}).
\end{eqnarray}
Using (\ref{multipole2}) and assuming axial symmetry of the nuclear shape, we
obtain \cite{sdhd} the following expressions for the intrinsic
electric multipole moments
\begin{eqnarray}
\label{multipole3}
Q_{10}&=&2D_{10}=e\frac{A}{2}(1-\eta ^2 )R_m (\frac{Z_1}{A_1}
-\frac{Z_2}{A_2}),\\
Q_{20}&=&e\frac{A}{4}(1-\eta ^2 )R^{2}_{m} ((1-\eta )\frac{Z_1}{A_1}
+(1+\eta )\frac{Z_2}{A_2})+Q_{20}(1)+Q_{20}(2),\\
Q_{30}&=&e\frac{A}{8}(1-\eta ^2 )R^{3}_{m} ((1-\eta )^2
\frac{Z_1}{A_1}-(1+\eta)^2
\frac{Z_2}{A_2})+\frac{3}{2}R_{m}((1-\eta )^2 Q_{20}(1)-(1+\eta )^2 Q_{20}(2)),
\end{eqnarray}
where the charge quadrupole moments of clusters $Q_{20}(i)\quad (i=1,2)$
are calculated with respect to their centers of mass.
Effective charges for electric dipole  and octupole transitions
are used in our calculations in order to take
the coupling of the mass--asymmetry mode
to the higher--lying giant dipole and octupole
excitations \cite{BM2} effectively into account, which are not present in the model.

The charge to mass ratios $Z_1/A_1$ and  $Z_2/A_2$ are
functions of $\eta$. For instance, for $|\eta |$=1 (mononucleus)
this ratio takes the values 0.3--0.4
for the nuclei considered in the paper. For the  $\alpha$--particle this
ratio is equal to 0.5. The results for the electric dipole moment
are sensitive to the dependence of $Z$/$A$ on $\eta$.
In the calculations we parameterize the
$Z_i/A_i$ ratio in the following way.
For $\eta_{\alpha}< |\eta|\le 1$,
the ratio $Z_2/A_2$ for the light cluster takes the same value
as for the  mononucleus.  For a smaller value of $|\eta |$, we set it
equal to 0.5 as for the $\alpha$--cluster.

\section{Results of calculations and discussion}

\subsection{Calculation procedure}

As was mentioned in Sect.~II.B, our consideration can be
restricted to
cluster configurations
near $|\eta |$=1. Then it is convenient to substitute the coordinate
$\eta$ by the following variable
\begin{eqnarray}
\label{newx}
x&=& \eta-1 \hspace{5pt} {\rm if} \hspace{5pt} \eta > 0, \nonumber \\
x&=& \eta+1 \hspace{5pt} {\rm if}  \hspace{5pt} \eta \le 0, \nonumber
\end{eqnarray}
and to use the following smooth parameterization
\begin{eqnarray}
U(x,I)=\sum_{k=0}^4 a_{2k}(I)x^{2k}
\label{poly_eq}
\end{eqnarray}
of the potential $U(\eta ,I)$ from Eq.~(\ref{pot_eq}).
This formula contains five parameters $a_{2k}(I)$.
If a minimum of the potential is located at $|x|$=$x_{\alpha}$ four parameters
are determined by the experimental ground state energy,
potential energies $U(x,I)$ for
$x=x_\alpha$, $x=x_{Li}$ and by the requirement that the potential
has a minimum at $|x|=x_{\alpha}$.
The fifth parameter $a_8$
is necessary to avoid a fall-off of the potential for
$|x| \ge x_{Li}$  because of the negative value of $a_6$
needed to describe correctly $U(x_{Li},I)$.
We take the minimal necessary
positive value of  $a_8$ to guarantee an increase of $U$ for
$|x| > x_{Li}$.
The ground state energy is obtained by solving the Schr\"odinger
equation. Since the ground state wave function is distributed over $x$,
the potential energy at $x=0$ is not equal to the experimental binding energy
of the mononucleus.
To reach
the correct value of the ground state energy $E_0(I=0)=0$,
we can vary the potential $U(x=0,I=0)$.
In the majority of cases this procedure leads to the value of
$U(x=0,I=0)$ close to $E_0(I=0)$.
The
variation of $B_\eta$ is also done in the case of light Ra isotopes
to obtain $E_0(I=0)$=0.
Besides the barrier height, which determines the stiffness of
the potential well at $x$=$x_\alpha$,
the  ground state energy $E_0(I=0)$ of
Eq.~(\ref{schr_eq}) depends also on the frequency of oscillations in
$x$.
This frequency is ruled by the value of the
inertia coefficient $B_\eta$.
If the minimum is located at $x=0$ ($|\eta |$=1), only
three parameters $a_0$, $a_2$ and $a_4$ in Eq.~(\ref{poly_eq})
are necessary. Potentials with other parameterizations show almost
no difference in the description of the parity splitting in the
majority of considered nuclei.

\subsection{Parity splitting}

With Eq.~(\ref{schr_eq})
we  first calculated
the parity splitting for the isotopes of Ra, Th, U, Pu, Ba, Ce and Nd
for different values of the angular momentum $I$.
The results of  calculations are shown in Tables 1--5.
As is seen from the Tables, they agree well
with the experimental data
\cite{exp,Cocks,Wiedenhover,Phillips,Ibbotson,Phillips1,Urban,Urban1,Ibbotson1,Wollersheim}.
The largest deviations
of the calculated values from the experimental ones are found
at low spins in some of the considered nuclei.
A good description of the experimental data, especially
of the variation
of the parity splitting with $A$ at low $I$ and of the value of the critical
angular momentum at which the parity splitting disappears,
means that
the dependence of the potential energy on
$\eta$ and $I$ for the considered nuclei
is described
correctly by the proposed cluster model.
The used value of the inertia coefficient $B_{\eta}$ is also important.

Of course, other effects related to degrees of freedom,
which are not included in the
model, like the alignment of the single particle momenta or interaction with other negative
parity bands with different $K$ quantum number can contribute as well. However, a
general agreement between the experimental data and the results of
calculations shows that the simple cluster model used in this paper gives a firm
ground for the consideration of the alternating parity bands.

In the considered nuclei the ground state energy level lies
near the top of the barrier in $\eta$, if exists, and
the
weight of the $\alpha-$cluster configuration (Fig.~2) estimated as that contribution
to the norm of the wave function which is located at
$|\eta|\le \eta_{\alpha}$ is about $5\times 10^{-2}$
for $^{226}$Ra, which is close to the calculated spectroscopic
factor \cite{Zam}. This means that our model is in qualitative agreement
with the known $\alpha-$decay widths of the nuclei considered.

The spectra of those considered nuclei whose potential energy has a minimum at
the alpha cluster configuration
can be well approximated by the following  analytical
expression
\begin{eqnarray}
E(I)&=&\frac{\hbar^2}{2J(I)}I[I+1],\quad {\rm if}\quad I\quad {\rm is \quad even}, \nonumber \\
E(I)&=&\frac{\hbar^2}{2J(I)}I[I+1]+\delta E(I),\quad {\rm if}\quad I\quad {\rm is \quad odd}.
\label{spectr_eq}
\end{eqnarray}
Here, the parity splitting $\delta E(I)$ is given as
\begin{eqnarray}
\delta E(I)=\frac{2E_1(I^\pi=1^-)}{1+\exp(b_0\sqrt {B_0I[I+1]})}
\label{spli_eq}
\end{eqnarray}
with
$$B_0=\frac{\hbar^2}{2}\left(\frac{1}{\Im (\eta=1)} -
\frac{1}{\Im (\eta=\eta_\alpha)}\right).$$\\
The quantity $B_0$ describes the  change of
the height of the barrier with spin $I$.
The moment of inertia in Eq.~(\ref{spectr_eq})
is given by the expression
\begin{eqnarray}
J(I)=w_m(I)\Im (\eta=1)+[1-w_m(I)]\Im (\eta=\eta_\alpha)
\end{eqnarray}
containing a weight function $w_m(I)$
\begin{eqnarray}
w_m(I)=\frac{w_m(I=0)}{1+b_1B_0I[I+1]},
\label{weight_eq}
\end{eqnarray}
which is the probability to find the mononucleus
component in the  wave function of the state with spin $I$
of the ground state band. Since $w_m(I)$ decreases with increasing angular momentum,
$J(I)$ increases with $I$ in agreement with an experimental tendency.
The quantity $w_\alpha(I)=1-w_m(I)$ gives the corresponding
probability of the $\alpha$-cluster component.
A qualitative  derivation of the above analytical formulae
is given in the Appendix.
The constants $\Im (|\eta|=1)=0.3\times \Im^r (|\eta|=1)$,
$w_m(I=0)$=0.93, $b_0=\pi$ MeV$^{-1/2}$ and  $b_1$=0.2 MeV$^{-1}$
were obtained by fitting the experimental spectra for the nuclei
considered (see Fig.~3).

These formulae clearly demonstrate that there are two important quantities
which predetermine a description of the spectra of the alternating
parity bands. They are $E_1(I^\pi=1^-)$, which is determined by the depth of the
minimum of the potential at $I=0$ and by the value of the mass parameter $B_{\eta}$, and $B_0$,
which determines the  angular momentum dependence of $w_m(I)$, i.e. of
$J(I)$ and $\delta E(I)$.

\subsection{$E\lambda$-transitions}

With the wave functions obtained, we have calculated the reduced matrix
elements of the electric multipole moments
$Q(E1)$, $Q(E2)$ and $Q(E3)$.
The effective charge for $E1$-transitions has
been taken to be equal to $e_1^{eff}=e(1+\chi)$
with an average state-independent
value of the $E1$ polarizability coefficient $\chi$=--0.7 \cite{BM2}.
This renormalization takes into account a
coupling of the mass--asymmetry mode to the giant dipole resonance
in a dinuclear system. In the case of the quadrupole transitions we did not
renormalized the charge $e^{eff}_2 =e$
although an effective charge of 1.35 $e$ describes the data for actinides better
as it is seen from the results of calculations.
For octupole transitions our cluster model Hamiltonian includes the octupole mode
responsible for description of the shape variation and deformation of the nuclear
surface. This is the low--frequency collective octupole mode.
However, high--frequency
isovector as well as isoscalar octupole modes are not presented in the model
Hamiltonian. For example, to simplify the consideration the charge asymmetry coordinate
is not independent dynamical one but is rigidly related in our model to the mass
asymmetry coordinate. The octupole transition operator is not exhausted by the term
produced by the low--frequency octupole degree of freedom and includes also a
contribution of the high--frequency octupole modes. For this reason
for the octupole transitions the effect of the
coupling of the low--frequency octupole mode to the high frequency mode should be taken
into account by
the octupole effective charge. The estimate of this effective charge is given in
\cite{BM2}.
The combined effect gives $\delta e_3^{(pol)}\approx (0.5+0.3 \tau_z )e$
\cite{BM2}.
So, we have taken the effective charge to be equal to
$e^{eff}_{3,proton}=1.2e$ for protons and $e^{eff}_{3,neutron}=0.8e$ for neutrons.

The results of these calculations are listed in
Tables 6, 7 and shown in Figs.~4--10.
The obtained values are in agreement with the
known experimental data for $Q^{exp}_\lambda$
\cite{butler,Cocks,Phillips,Ibbotson,Phillips1,Urban,Urban1,Ibbotson1,Mach,Pitz,Wollersheim,Raman}.
Only in $^{224}$Ra and $^{146}$Ba (for $I=7$) the calculated
values of $D_{10}$ are larger by factor of four than the experimental $D_{10}$.
In Th the isotopic dependence of the dipole moment is
well reproduced.
The higher multipole moments  are in agreement
with the calculations of Ref.~\cite{mollernix}.
Taking into account the collective character of our model and
the absence of the parameters to fit the data, the description
of the experimental data is rather good.
It should be also noted that the experimental data on the dipole moment have
some uncertainties.

The angular momentum dependence of the reduced matrix elements of the
electric dipole operator is presented in
Figs.~4 and 7 for $^{226}$Ra and $^{148}$Nd,
respectively.
The calculations qualitatively  reproduce the angular
momentum dependence of the experimental matrix elements
\cite{Ibbotson,Wollersheim}.
The same is true for the reduced matrix elements
of the electric quadrupole and octupole operators (Figs.~5,6 and 8,9).

Fig.~10 illustrates the angular momentum dependence of the
calculated intrinsic transition   quadrupole
moment. It is interesting that the cluster model shows an increase
of the quadrupole moment with angular momentum in the transitional nucleus
$^{226}$Ra and a constant dependence in the well deformed isotope $^{238}$U.
Staggering shown in Fig.~10  for both $^{226}$Ra and  $^{238}$U nuclei is
explained by the higher weight of the $\alpha$--cluster component in the wave
functions of odd $I$ states (see Fig.~2).
This cluster configuration has larger quadrupole and octupole deformations.

The calculated results for the E3--reduced matrix elements in $^{148}$Nd exceed
the experimental data for transitions to the ground band \cite{Ibbotson}.
This can be explained as follows.
The experimental E3 matrix elements connecting
the negative parity states of $^{148}$Nd
to the $\beta$ band are unexpectedly large, about  70$\%$ of
the matrix elements within the ground band \cite{Cline}.
This shows a considerable fractionation of the E3 strength among the $K$=0
bands. In our model the $\beta$ degree of freedom is absent
and all the E3 strength is
concentrated in the transitions to the ground band.
The nucleus $^{148}$Nd is transitional
in its collective properties between spherical and
deformed nuclei and the $\beta$ anharmonicity
is quite large. This explains a fractionation of the E3
strength between the ground and
the $\beta$ bands.
The summed E3 strength for $^{148}$Nd corresponds to the
intrinsic transitional
octupole moment of $\quad\sim 2000 e$ fm$^3$
(instead of $\quad\sim 1500 e$ fm$^3$
for the transitions in the ground band)
which agrees with the calculated value.

\section{Summary}

We suggest a cluster interpretation
of the properties of the
alternating parity bands in heavy nuclei
assuming collective oscillations in
mass asymmetry degree of freedom.
The existing experimental data on the
angular momentum  dependence of parity splitting
and on  multipole transition  moments are quite well reproduced.
This supports the idea that cluster type states exist in heavy nuclei.
The characteristics of the Hamiltonian used in the calculations were
determined by investigating a completely different
phenomenon, namely,
heavy ion reactions at low energies.
Due to this fact, a predictive power of the
suggested model is quite high.
The proposed analytical expression (\ref{spectr_eq})
for $E(I)$ can be applied to estimate the position of
the low--lying states which are not yet measured.
The calculated staggering behavior of alpha--clusterization
can be verified by measuring the
angular momentum dependence of the width of the alpha--decay.

\section{Acknowledgment}

T.M.S. and R.V.J. are grateful to
BMBF and DFG (Bonn), respectively, for support.
This work was supported in part by Volkswagen-Stiftung (Hannover) and
RFBR (Moscow). The support of  STCU(Uzb--45), SCST and UFBR (Tashkent) is
acknowledged as well.

\appendix
\section{}
Let us assume that the barrier at $x$=0 separates two minima of the potential
(\ref{pot_eq}): the minimum
with reflection asymmetric deformation
(minimum of the potential at $x=x_\alpha$)
and its mirror image. The nonzero penetration through this barrier
lowers the  energy of the levels with even $I$ with respect to
the energy of the levels with odd $I$. With increasing spin $I$ the
barrier between two minima, which is equal to
$U_b(I)=U_b(I=0)+B_0I[I+1]$, becomes higher and
the penetration probability goes to zero.
According to standard WKB-analysis (with higher order  corrections)
the transmission probability (per tunneling event) for the potential
barrier described by the inverted
oscillator with frequency $\hbar\omega_b(I)$
for the energy $\hbar\omega_m(I)/2$ above the potential minimum is given by
\cite{Landau}
\begin{eqnarray}
P(I)=\frac{1}{1+\exp(2\pi\frac{U_b(I)-\hbar\omega_m(I)/2}{\hbar\omega_b(I)})},
\label{trans_eq}
\end{eqnarray}
and  $\hbar\omega_b(I)\sim\sqrt {U_b(I)}$ at  fixed $x_\alpha$.
Here, $\omega_m(I)$ is the frequency of the harmonic
oscillator which approximates the potential $U$
around the $\alpha$-cluster minimum.
Within the semiclassical
approximation  one can neglect the $I$ dependence of $\omega_m$ taking
$\omega_m (I)\approx\omega_m (I=0)$.
This frequency then essentially determines
the rate $\omega_m/\pi$ at which the wave packet strikes
the barrier, so that the effective coherent tunneling frequency, i.e.
the shift of the negative parity states with respect to the positive parity
ones, is given by
\begin{eqnarray}
\delta E(I)=\frac{\hbar\omega_m}{\pi}P(I).
\end{eqnarray}
Assuming that $(U_b(I=0)-\frac{1}{2}\hbar\omega_m )$ is small compared
to $\hbar\omega_b (0)$ (this case is realized in nuclei considered in the
paper), we obtain  $\hbar\omega_m$/2$\pi$ =$E(I^{\pi}=1^- )$.
For those values of $I$ at which $U_b (I=0)$ is much smaller than $B_0 I(I+1)$
we obtain Eq.~(\ref{spli_eq})
with the fitting parameter $b_0$.
However, we found numerically that Eq.~(\ref{spli_eq}) works quite well also
at low $I$.

The weight $w_m(I)$ of the mononucleus
component in the wave function of the state with spin $I$ can be expressed
through the ratio of characteristic times $\tau_m(I)$ and
$\tau_b(I)$ which a system spends in the minima and at the barrier,
respectively
\begin{eqnarray}
w_m(I)=\frac{\tau_b(I)}{\tau_b(I) + \tau_m(I)}.
\label{weight1_eq}
\end{eqnarray}
The mononucleus configuration is located at the top of the barrier.
We neglect
below the $I$ dependence of the $\tau_m(I)$. Denote by $\tau_b(0)$
the value of $\tau_b(I)$ at $I$=0. At very high angular momentum when the
barrier height is mainly determined by the rotational energy and is equal to
$B_0 I(I+1)$ the value of $\tau_b(I)$ can be determined from the time--energy uncertainty relation
\begin{eqnarray}
\tau_b (I\gg 1)=\frac{\hbar}{B_0 I(I+1)}.
\label{weight2_eq}
\end{eqnarray}
To combine the two limits at $I$=0 and for $I\gg 1$, we use the following expression
\begin{eqnarray}
\tau_b(I)= \frac{ \hbar }{ \hbar /\tau_b(0) + U_b(I)-\hbar\omega_m / 2 }
               = \frac{ \tau_b(0) }{ 1 + \tau_b(0)B_0I[I+1] / \hbar } .\\
\label{weight3_eq}
\end{eqnarray}
Substituting this result into (\ref{weight1_eq}), we obtain
\begin{eqnarray}
w_m(I)=  \frac{ \tau_b(0) / ( \tau_b(0) + \tau_m ) }
                        { 1 + \frac{ \tau_b(0)\tau_m }{ \hbar
(\tau_b(0) + \tau_m) } B_0I[I+1] }.
\label{weight4_eq}
\end{eqnarray}
The last expression can be rewritten as
\begin{eqnarray}
w_m(I)=\frac{w_m(0)}{1+b_1 B_0 I(I+1)},
\label{weight5_eq}
\end{eqnarray}
where $w_m(0)=\tau_b(0)/(\tau_b(0)+\tau_m)$ and
$b_1 =\frac{1}{\hbar}\tau_b(0)\tau_m /(\tau_m + \tau_b(0) )$.

\begin{figure}
\caption{Potential energy (full circles) of the $\alpha$--cluster
configuration $U(\eta_{\alpha})\equiv U(\eta_{\alpha},I=0)$
as a function of the neutron number in  Ba isotopes.}
\label{1_fig}
\end{figure}
\begin{figure}
\caption{Calculated probability of the $\alpha-$cluster component
in the wave function of the state with spin $I$ of the alternating
parity band for $^{224,226}$Ra and $^{226,232}$Th.}
\label{2_fig}
\end{figure}
\begin{figure}
\caption{Comparison of experimental (solid points) and
calculated with Eqs. (\protect\ref{spectr_eq})-(\protect\ref{weight_eq})
(solid lines) energies of states of the
alternating parity bands in
$^{220,224}$Ra and $^{222}$Th. The fitting parameters are the same
for all nuclei (see text). Experimental data are taken
from \protect\cite{exp,Cocks}.}
\label{3_fig}
\end{figure}
\begin{figure}
\caption{Angular momentum dependence of the calculated reduced matrix elements
of the electric dipole operator (solid curve) in $^{226}$Ra. The experimental
data (squares) are taken from \protect\cite{Wollersheim}}
\label{4_fig}
\end{figure}
\begin{figure}
\caption{The same as in Fig. 4, but for the quadrupole operator}
\label{5_fig}
\end{figure}
\begin{figure}
\caption{The same as in Fig. 4, but for the octupole operator}
\label{6_fig}
\end{figure}
\begin{figure}
\caption{Angular momentum dependence of the calculated reduced matrix elements
of the electric dipole operator (solid curve) in $^{148}$Nd. The experimental
data (squares) are taken from \protect\cite{Ibbotson}.}
\label{7_fig}
\end{figure}
\begin{figure}
\caption{The same as in Fig. 7, but for the quadrupole operator}
\label{8_fig}
\end{figure}
\begin{figure}
\caption{The same as in Fig. 7, but for the octupole operator}
\label{9_fig}
\end{figure}
\begin{figure}
\caption{Angular momentum dependence of  calculated intrinsic
quadrupole transition moments in $^{226}$Ra and $^{238}$U.}
\label{10_fig}
\end{figure}

\begin{table}[here]
\caption{Comparison of experimental ($E_{exp}$) and
calculated ($E_{calc}$) energies of states of the
alternating parity bands in
$^{232-222}$Th. Energies are
given in keV. Experimental data are taken from \protect\cite{exp,Cocks}.}
\begin{tabular}{|l|lr|lr|lr|lr|lr|lr|}
 & $^{232}$Th & & $^{230}$Th & & $^{228}$Th & &  $^{226}$Th & & $^{224}$Th & & $^{222}$Th &\\
\hline
I$^\pi$ &  $E_{exp}$ & $E_{calc}$ & $E_{exp}$ &$E_{calc}$ &  $E_{exp}$
& $E_{calc}$ &
 $E_{exp}$ & $E_{calc}$ &  $E_{exp}$ & $E_{calc}$ &  $E_{exp}$ & $E_{calc}$ \\
\hline
1$^-$ & 714 & 693 & 508 & 485 & 328 & 350 & 230 & 254 & 251 & 204 & 250 &195 \\
2$^+$ & 49 & 49 & 53 & 53 & 58 & 58 & 72 & 72 & 98 & 98 & 183 & 183 \\
3$^-$ & 774 & 761 & 572 & 557 & 396 & 423 & 308 & 340 & 305 & 311 & 467 & 366 \\
4$^+$ & 162 & 160 & 174 & 172 & 187 & 177 & 226 & 238 & 284 & 296 & 440 & 461 \\
5$^-$ & 884 & 882 & 687 & 684 & 519 & 549 & 451 & 490 & 465 & 494 & 651 & 616 \\
6$^+$ & 333 & 330 & 357 & 354 & 378 & 391 & 447 & 475 & 535 & 563 & 750 & 760 \\
7$^-$ & 1043 & 1051 & 852 & 859 & 695 & 748 & 658 & 698 & 700 & 739 & 924 & 920 \\
8$^+$ & 557 & 553 & 594 & 589 & 623 &634 & 722 & 761 & 834 & 868 & 1094 & 1077 \\
9$^-$ & 1249 & 1263 & 1065 & 1075 & 921 & 971 & 923 & 958 &  998 & 1036 & 1255 & 1258 \\
10$^+$& 827 & 822 & 880 & 869 & 912 & 919 & 1040 & 1079 & 1174 & 1202 & 1461 & 1430 \\
11$^-$& 1499 & 1511 & 1322 & 1326 & 1190 & 1229 & 1238 & 1263 & 1347 & 1384 & 1623 & 1624 \\
12$^+$& 1137 & 1130 & 1208 & 1215 & 1239 & 1235 & 1395 & 1424 & 1550 & 1564 & 1851 & 1815\\
13$^-$& 1785 & 1792 & 1615 & 1629 & 1497 & 1517 & 1596 & 1609 & 1739 & 1772 & 2016 & 2019 \\
14$^+$& 1482 & 1470 & 1573 & 1565 & 1605 & 1572 & 1781 & 1796 & 1959 & 1966 & 2260 & 2226 \\
15$^-$& 2101 & 2099 & 1946 & 1941 & 1838 & 1823 & 1989 & 2002 & 2165 & 2194 & 2432 & 2450 \\
16$^+$& 1858 & 1841 & 1971 & 1935 & 1993 & 1918 & 2196 & 2200 & 2398 & 2405 & 2688 & 2663 \\
17$^-$& 2445 & 2449 & 2310 & 2274 & 2209 & 2154 & 2413 & 2429 & 2620 & 2651& 2873 & 2906 \\
18$^+$& 2262 & 2229 & 2398 & 2318 & 2406 & 2281 & 2635 & 2640 & 2864 & 2880 & 3134 & 3128 \\
19$^-$& 2813 & 2794 & 2703 & 2624 &      &      & 2861 & 2890 & & &  3341 & 3380 \\
20$^+$& 2691 & 2633 & 2850 & 2709 &      &      & 3097 & 3115 & & & 3596 & 3621 \\
\end{tabular}
\end{table}

\newpage
\begin{table}[here]
\caption{Comparison of experimental ($E_{exp}$) and calculated ($E_{calc}$)
energies of states of the alternating parity bands in
$^{220-226}$Ra and $^{240,242}$Pu . Energies are
given in keV. Experimental data are taken from \protect\cite{exp,Cocks,Wiedenhover}.
For $^{220,222}$Ra,
the parameter $c_2$ was adjusted to the 6$^+$ state.}
\begin{tabular}{|l|lr|lr|lr|lr|lr|lr|}
 & $^{226}$Ra & & $^{224}$Ra & & $^{222}$Ra & &  $^{220}$Ra & & $^{242}$Pu &
 &$^{240}$Pu & \\
\hline
I$^\pi$ &  $E_{exp}$ & $E_{calc}$ & $E_{exp}$ & $E_{calc}$ &  $E_{exp}$
& $E_{calc}$ & $E_{exp}$ & $E_{calc}$ &  $E_{exp}$
& $E_{calc}$ & $E_{exp}$ & $E_{calc}$ \\ \hline
1$^-$ & 254 & 254 & 216 & 193 & 242 & 224 & 413 & 385 & 781 & 778 & 597 & 597  \\
2$^+$ & 68  & 68  & 85  & 85  & 111 & 96  & 179 & 125 &  45 &  45 &  43 &  43  \\
3$^-$ & 322 & 327 & 291 & 282 & 317 & 324 & 474 & 509 & 832 & 843 & 649 & 659 \\
4$^+$ & 212 & 206 & 251 & 253 & 302 & 287 & 410 & 375 & 147 & 146 & 142 & 142 \\
5$^-$ & 447 & 455 & 433 & 434 & 474 & 486 & 635 & 709 & 927 & 958 & 742 & 774\\
6$^+$ & 417 & 414 & 480 & 482 & 550 & 550 & 688 & 688 & 306 & 304 & 294 & 295 \\
7$^-$ & 627 & 635 & 641 & 642 & 703 & 728 & 873 & 962 &     &1122 &     & 945 \\
8$^+$ & 670 & 668 & 756 & 747 & 843 & 843 & 1001& 1016& 518 & 514 & 498 & 499 \\
9$^-$ & 858 & 862 & 907 & 901 & 992 & 1014& 1164& 1252&     &1329 &     &1145  \\
10$^+$& 960 & 954 & 1069& 1046& 1173& 1166& 1343& 1356& 779 & 773 & 748 & 750 \\
11$^-$& 1134& 1134 &1222& 1215& 1331& 1346& 1496& 1568&     &1578 &     &1392 \\
12$^+$& 1282& 1270 &1415& 1378& 1537& 1525& 1711& 1706&1084 &1077 &1042 &1044 \\
13$^-$& 1448& 1449 &1578& 1558& 1710& 1722& 1864& 1904&     &1863 &     &1677 \\
14$^+$& 1629& 1621 &1789& 1745& 1933& 1924& 2106& 2067&1431 &1421 &1376 &1377 \\
15$^-$& 1797& 1810 &1970& 1944& 2125& 2140& 2263& 2257&     &2181 &     &1994   \\
16$^+$& 1999& 2003 &2189& 2153& 2359& 2366&     &     &1816 &1800 &     &   \\
17$^-$& 2175& 2220 &2389& 2372& 2570& 2602&     &     &     &2526 &     &   \\
18$^+$&     &      &    &     &     &     &     &     &2236 &2210 &     &\\
19$^-$&     &      &    &     &     &     &     &     &     &2894 &     &\\
20$^+$&     &      &    &     &     &     &     &     &2686 &2646 &     &\\
\end{tabular}
\end{table}

\newpage
\begin{table}[here]
\caption{Comparison of experimental ($E_{exp}$) and calculated ($E_{calc}$)
energies of states of the  alternating parity bands in
$^{238-232}$U. Energies are given in keV.
Experimental data are taken from \protect\cite{exp}.}
\begin{tabular}{|l|lr|lr|lr|lr|}
 & $^{238}$U & & $^{236}$U & & $^{234}$U & &  $^{232}$U &\\
\hline
I$^\pi$ &  $E_{exp}$ & $E_{calc}$ & $E_{exp}$ & $E_{calc}$ &  $E_{exp}$
& $E_{calc}$ &  $E_{exp}$ & $E_{calc}$  \\ \hline
1$^-$ & 680 & 675 & 688 & 644 & 786 & 778 & 563 & 583  \\
2$^+$ & 45  & 45  & 45  &  45 & 44  & 44  &  48 & 48   \\
3$^-$ & 732 & 744 & 744 & 713 & 849 & 846 & 629 & 653  \\
4$^+$ & 148 & 156 & 150 & 154 & 143 & 155 & 157 & 158  \\
5$^-$ & 827 & 863 & 848 & 831 & 963 & 963 & 747 & 774  \\
6$^+$ & 307 & 316 & 310 & 313 & 296 & 314 & 323 & 320  \\
7$^-$ & 966 & 1025 & 1000 & 992 & 1125 & 1122 & 915 & 938  \\
8$^+$ & 518 & 520 & 522 & 516 & 497 & 517 & 541 & 527 \\
9$^-$ & 1150 & 1222 & 1199 & 1189 & 1336 & 1316 & 1131 & 1138  \\
10$^+$& 776 & 759 & 782 & 753 & 741 & 754 & 806 & 768 \\
11$^-$& 1378 & 1448 &  &  &  &  & 1391 & 1366 \\
12$^+$& 1077 & 1025 &  &  &  &  & 1112 & 1036 \\
\end{tabular}
\end{table}

\newpage

\begin{table}[here]

\caption
{Comparison of experimental ($E_{exp}$) and calculated ($E_{calc}$)
energies of states of the ground state alternating parity bands in
$^{144-148}$Ba. Energies are given in keV.
Experimental data are taken from \protect\cite{exp,Phillips}.
The parameter $c_2$ was adjusted to the 6$^+$ state.}
\begin{tabular}{|l|lr|lr|lr|}
 & $^{148}$Ba & & $^{146}$Ba & & $^{144}$Ba &\\
\hline
I$^\pi$ &  $E_{exp}$ & $E_{calc}$ & $E_{exp}$ & $E_{calc}$ &  $E_{exp}$
& $E_{calc}$  \\
\hline
1$^-$ &      & 623 & 739 & 664  & 759&   607  \\
2$^+$ & 142  & 124 & 181 & 143  & 199 &  157    \\
3$^-$ & 775  & 771 & 821 & 818  & 838 & 763   \\
4$^+$ & 423  & 400 & 514 & 469  & 530 &  505    \\
5$^-$ & 963  & 1018&1025 & 1078 & 1039&  1026   \\
6$^+$ & 808  & 808 & 958 & 958  & 961 &  961   \\
7$^-$ & 1256 & 1342&1349 & 1424 & 1355&  1375  \\
8$^+$ & 1265 & 1273&1483 & 1491 & 1471&  1496  \\
9$^-$ & 1645 & 1731&1778 & 1841 & 1772&  1796   \\
10$^+$& 1768 & 1788&2052 & 2028 & 2044&  2005\\
11$^-$& 2117 & 2181&2293 & 2323 & 2278&  2285  \\
12$^+$& 2304 & 2327&2632 & 2574 & 2667&  2546  \\
13$^-$&      &     &2877 & 2871 & 2863&  2843  \\
14$^+$&      &     &3193 & 3166 & 3321&  3146  \\
15$^-$&      &     &3524 & 3489 & 3519&  3473  \\
16$^+$&      &     &3737 & 3823 & 3992&  3815  \\
17$^-$&      &     &     &      & 4242&  4179  \\
\end{tabular}
\end{table}

\newpage
\begin{table}[here]
\caption
{Comparison of experimental ($E_{exp}$) and calculated ($E_{calc}$)
energies of states of the ground state alternating parity bands in
$^{146,148}$Ce and $^{146,148}$Nd isotopes. Energies are given in keV.
Experimental data are taken from \protect\cite{exp,Ibbotson,Phillips1,Urban,Urban1,Ibbotson1}.
The parameter $c_2$ was adjusted to the 6$^+$ state.}
\begin{tabular}{|l|lr|lr|lr|lr|}
 & $^{148}$Ce & & $^{146}$Ce & & $^{148}$Nd && $^{146}$Nd &\\
\hline
I$^\pi$ &  $E_{exp}$ & $E_{calc}$ & $E_{exp}$ & $E_{calc}$ &  $E_{exp}$
& $E_{calc}$ &  $E_{exp}$
& $E_{calc}$ \\
\hline
1$^-$ & 760  & 714 & 925  & 776  & 1023& 734  &     & 896  \\
2$^+$ & 159  & 134 & 259  & 195  & 302 & 279  & 454 & 327  \\
3$^-$ & 841  & 851 & 961  & 956  & 999 & 943  & 1190& 1202 \\
4$^+$ & 453  & 424 & 668  & 614  & 752 & 776  & 1042& 993  \\
5$^-$ &      & 1084& 1183 & 1259 & 1242& 1261 & 1518& 1684  \\
6$^+$ & 840  & 840 & 1171 & 1171 & 1280& 1280 & 1780& 1780  \\
7$^-$ &      & 1400& 1550 & 1660 & 1645& 1647 & 2029& 2264  \\
8$^+$ & 1290 & 1289& 1737 & 1756 & 1856& 1788 & 2594& 2510  \\
9$^-$ &      & 1790& 2019 & 2138 & 2132& 2084 & 2706& 2889   \\
10$^+$& 1792 & 1793& 2552 & 2345 & 2472& 2286 & 3320& 3195   \\
11$^-$&      & 2246& 2562 & 2681 & 2677& 2573 & 3501& 3544  \\
12$^+$& 2328 & 2334& 3013 & 2953 & 3107& 2819 & 3998& 3879  \\
13$^-$&      & 2769& 3163 & 3286 & 3265& 3120 & 4295& 4235  \\
14$^+$& 2888 & 2919&      & 3603 &     &      & 4694& 4594  \\
15$^-$&      & 3358& 3827 & 3954 &     &      & 5058& 4970  \\
16$^+$& 3464 & 3554&      &      &     &      & 5461& 5356   \\
17$^-$&      & 4013&      &      &     &      &     &\\
18$^+$& 4065 & 4243&      &      &     &      &     & \\
19$^-$&      & 4735&      &      &     &      &     &\\
20$^+$& 4685 & 4983&      &      &     &      &     &\\
\end{tabular}
\end{table}
\newpage
\begin{table}[here]
\caption
{Calculated and experimental intrinsic multipole transition moments. The
values of the dipole moment $D_{10}$ are given for
those values of the nuclear spin
$I$ for which there are experimental data. These values of $I$
are shown in the second
column. The experimental data are taken from
\protect\cite{butler,exp,Cocks,Wiedenhover,Wollersheim,Raman}.}
\begin{tabular}{|l|c|c|c|c|c|c|}
Nucleus & $D_{10}$ & $D_{10}$ & $Q_{20}(0^+\to 2^+)$ &
$Q_{20}(0^+\to 2^+)$ & $Q_{30}(0^+\to 3^-)$ & $Q_{30}(0^+\to 3^-)$\\
 & ($e$ fm) & ($e$ fm) & ($e$ fm$^2$) &  ($e$
 fm$^2$) &  ($e$  fm$^3$) &  ($e$ fm$^3$) \\
 & calc. & exp. & calc. & exp. & calc. & exp. \\
\tableline
$^{220}$Ra & 0.28 ($I$=7)  & 0.27 &397  & 558 &3167&          \\
$^{222}$Ra & 0.30 ($I$=7) & 0.27 & 395 & 675 & 3064 &         \\
$^{224}$Ra & 0.133 ($I$=3) & 0.028 & 510 & 633 & 2889 &         \\
$^{226}$Ra & 0.111 ($I$=1) & 0.06--0.10 & 574 & 718 & 2611 & 2861 \\
$^{222}$Th & 0.29 ($I$=6) & 0.38  & 397  & 548 & 3632 &     \\
$^{224}$Th & 0.312 ($I$=10) & 0.52 & 495 &  & 2985 &     \\
$^{226}$Th & 0.223 ($I$=8) & 0.30 & 561 & 830  & 2672 &     \\
$^{228}$Th & 0.151 ($I$=8) & 0.12 & 653 & 843  & 2255 &     \\
$^{230}$Th & 0.054 ($I$=6) & 0.04 & 666 & 899  & 1935 & 2144    \\
$^{232}$Th & 0.007 ($I$=1) &      & 719 & 966  & 1616 & 1969    \\
$^{234}$U & 0.004 ($I$=1) &  & 758 & 1035  & 1541 & 1895    \\
$^{236}$U & 0.004 ($I$=1) &  & 786 & 1080  & 1433 & 1951    \\
$^{238}$U & 0.004 ($I$=1) &  & 818 & 1102  & 1417 & 2041    \\
\end{tabular}
\end{table}

\newpage
\begin{table}[here]
\caption
{Calculated and experimental intrinsic multipole transition moments for Ba,
Ce and Nd isotopes. The
values of the dipole moment $D_{10}$ are given for
those values of the nuclear spin
$I$ for which there are experimental data. These values of $I$
are shown in the second
column. The experimental data are taken from
\protect\cite{Phillips,Ibbotson,Phillips1,Urban,Urban1,Ibbotson1,Mach,Pitz}.}
\begin{tabular}{|l|c|c|c|c|c|c|}
Nucleus & $D_{10}$ & $D_{10}$ & $Q_{20}(0^+\to 2^+)$ &
$Q_{20}(0^+\to 2^+)$ & $Q_{30}(0^+\to 3^-)$ & $Q_{30}(0^+\to 3^-)$\\
 & ($e$ fm) & ($e$ fm) & ($e$ fm$^2$) &  ( $e$
 fm$^2$) &  ( $e$  fm$^3$) &  ($e$ fm$^3$) \\
 & calc. & exp. & calc. & exp. & calc. & exp. \\
\tableline
$^{144}$Ba & 0.194 ($I$=7)  & 0.071(10) &250  & 321 &1295&          \\
           & 0.209 ($I$=8) & 0.14(3)    &     &     &    &          \\
$^{146}$Ba & 0.055 ($I$=3) & 0.06(4) & 286 & 368 & 1147 &         \\
           & 0.170 ($I$=7) & 0.037(3)    &     &     &    &          \\
$^{148}$Ba & 0.095 ($I$=7) &         & 306 &     & 938    &         \\
$^{146}$Ce & 0.121 ($I$=7) & 0.11(2) & 313 & 305 & 1669&  \\
$^{146}$Ce & 0.160 ($I$=11) & 0.20(2) &  &  & &  \\
$^{148}$Ce & 0.152 ($I$=7) &     & 364  & 436 & 1771 &     \\
$^{146}$Nd & 0.071 ($I$=11) & 0.17(2) & 264 & 276 & 1627 &     \\
$^{148}$Nd & 0.115($I$=1) & 0.24(6) & 370 & 400  & 2161 & 1500    \\
           & 0.222 ($I$=8) & 0.24(3)    &     &     &    &          \\
\end{tabular}
\end{table}


\begin{thebibliography}{99}
\bibitem{ahmad} I. Ahmad and P.A. Butler,
Ann. Rev. Nucl. Part. Sci. {\bf 43}, 71 (1993).
\bibitem{butler} P.A. Butler and W. Nazarewicz,
Rev. Mod. Phys. {\bf 68}, 350 (1996).
\bibitem{Moeller} P.~M\"oller and S.G.~Nilsson,
Phys. Lett. B {\bf 31}, 283 (1970).
\bibitem{Leander} G.A.~Leander, R.K.~Sheline, P.~M\"oller, P.~Olanders,
  I.~Ragnarsson, and A.J.~Sirk, Nucl. Phys. A {\bf 388}, 452 (1982).
\bibitem{Sobic} A.~Sobiczewski and B\"oning,
Acta Phys. Pol. B {\bf 18} 393 (1987).
\bibitem{Robledo} L.M.~Robledo, J.L.Egido, J.F.~Berger, and M.~Girod,
  Phys. Lett. B {\bf 187}, 223 (1987).
\bibitem{Iachello} F.Iachello and A.D.Jackson, Phys. Lett. B {\bf 108},
151 (1982).
\bibitem{Gai} M.Gai {\it et al.}, Phys. Rev. Lett. {\bf 51}, 646
(1983).
\bibitem{Daley} H.Daley and F.Iachello, Phys. Lett. B {\bf 131}, 281 (1983).
\bibitem{Daley1} H.Daley and J.Barret, Nucl. Phys. A {\bf 449}, 256 (1986).
\bibitem{Wilder}K.~Wildermuth and Y.C. Tang,  {\it A Unified Theory of the
Nucleus} (Vieweg, Braunschweig, 1977).
\bibitem{Bromley} D.A.~Bromley, Sci. Am. {\bf 239}, 58 (1978).
\bibitem {Cindro} N.Cindro and W.Greiner,  J. Phys. G {\bf 9}, L175 (1983).
\bibitem{Zamfir} N.V.Zamfir and D.Kuznezov, Phys. Rev. C {\bf 63}, 054306 (1985).
\bibitem{Zamfir1} N.V.Zamfir, {\it International Symposium on
Nuclear Structure Physics}, eds. R.F.Casten {\it et al.}
(World Scientific, Singapore, 2001).
\bibitem{Buck} B.Buck, A.C.Merchant, and S.M.Perez,
Phys. Rev. Lett. {\bf 76}, 380 (1996).
\bibitem{Buck1} B.Buck, A.C.Merchant, and S.M.Perez, Phys. Rev. C {\bf 58},
2049 (1998).
\bibitem{Buck2} B.Buck, A.C.Merchant, and S.M.Perez,
Phys. Rev. C {\bf 59}, 750 (1999).
\bibitem{Buck3} B.Buck, A.C.Merchant, and S.M.Perez,
Phys. Rev. C {\bf  61}, 024314 (2000).
\bibitem{Rae} W.D.M.~Rae, Int.J.Mod.Phys. A {\bf 3}, 1343 (1988).
\bibitem{Freer1} M.~Freer, R.R.~Betts, and A.H.~Wuosmaa,
Nucl. Phys. A {\bf 587}, 36 (1995).
\bibitem{Freer2} M.~Freer and A.C.~Merchant, J. Phys. G {\bf 23}, 261 (1997).
\bibitem{Horiuchi1} H.~Horiuchi, Nucl.Phys. A {\bf 552}, 257c (1991).
\bibitem{Horiuchi} H.~Horiuchi, and Y.K.~Kanada--En'yo, Nucl.Phys. A
{\bf 616}, 394 (1997);\\
Y.K.~Kanada--En'yo, Phys. Rev. Lett. {\bf 81}, 5291 (1998).
\bibitem{Zam} Yu.S.Zamiatin {\it et al.}, Phys. Part. Nucl. {\bf 21},
 537 (1990).
\bibitem{Pashkevich} V.V. Pashkevich {\it et al.},
Nucl. Phys. A {\bf 624}, 140 (1997).
\bibitem{Aberg} S.~Aberg and L.--O.Jonsson,  Z. Phys. A {\bf 349}, 205 (1994).
\bibitem{sdhd}T.M.~Shneidman, G.G.~Adamian, N.V.~Antonenko,
S.P.~Ivanova, and W.~Scheid,
Nucl.~Phys. A {\bf 671}, 119 (2000).
\bibitem{Jolos1} R.V.~Jolos, P. von Brentano, and F.~D\"onau, J. Phys. G
{\bf 19}, L151 (1993).
\bibitem{Jolos2} R.V.~Jolos and P. von Brentano, Phys.~Rev. C {\bf 49},
R2301 (1994); Nucl.~Phys. A {\bf 587}, 377 (1995).
\bibitem{Shneidman} T.M.~Shneidman, G.G.~Adamian, N.V.~Antonenko, R.V.~Jolos,
  and W.~Scheid, Phys. Lett. B {\bf 526}, 322 (2002).
\bibitem{volkov} V.V.Volkov, Phys. Rep. {\bf 44}, 93 (1978).
\bibitem{obzor} G.G.Adamian, A.K.Nasirov, N.V.Antonenko, and R.V.Jolos,
Phys. Part. Nucl. {\bf 25}, 583 (1994).
\bibitem{isot}G.G.Adamian,~N.V.Antonenko, and W.~Scheid,
Nucl.~Phys.~A {\bf 678}, 24 (2000).
\bibitem{Greiner} W.Greiner, J.Y.Park, and W.Scheid,
{\it Nuclear Molecules} (World Scientific, Singapore, 1995).
\bibitem{poten} G.G.Adamian {\it et al.}, Int. J.  Mod. Phys. E {\bf 5},
191 (1996).
\bibitem{Migdal} A.B.Migdal,  {\it Theory of Finite Fermi Systems and
Applications to Atomic Nuclei} (John Wiley $\&$ Sons, NY, 1967).
\bibitem{mollernix} P.M\"oller {\it et al.}, At. Data and Nucl. Tables {\bf
59}, 185 (1995).
\bibitem{mass} G.G.Adamian, N.V.Antonenko, and R.V.Jolos,
Nucl. Phys. A {\bf 584}, 205 (1995).
\bibitem{BM2} A.~Bohr and B.R.~Mottelson, {\it Nuclear Structure}, Vol. II
 (Benjamin, New York, 1975).
\bibitem{exp} http://www.nndc.bnl.gov/nndc/ensdf/
\bibitem{Cocks} J.F.C.~Cocks {\it et al.}, Nucl. Phys. A {\bf 645}, 61 (1999).
\bibitem{Wiedenhover} I.~Wiedenhover {\it et al.}, Phys. Rev. Lett.
{\bf 83}, 2143 (1999).
\bibitem{Phillips} W.R.~Phillips, I.~Ahmad, H.~Emling {\it et al.},
Phys. Rev. Lett. {\bf 57}, 3257 (1986).
\bibitem{Ibbotson} R.~Ibbotson, C.A.~White,
T.~Czosnyka, P.A.~Butler {\it et al.},
Phys. Rev. Lett. {\bf 71}, 1990 (1993).
\bibitem{Phillips1} W.R.~Phillips, R.V.F.~Janssens,
I.~Ahmad {\it et al.}, Phys.Lett. B {\bf 212}, 402 (1988).
\bibitem{Urban} W.~Urban, R.M.~Lieder, W.~Gast {\it et al.},
Phys. Lett. B {\bf 200}, 424 (1988).
\bibitem{Urban1} W.~Urban, R.M.~Lieder, J.C.~Bacelar  {\it et al.},
Phys.Lett. B {\bf 258}, 293 (1991).
\bibitem{Ibbotson1} R.~Ibbotson,
B.~Kotlinski, D.~Cline {\it et al.}, Nucl.Phys. A {\bf 530}, 199 (1991).
\bibitem{Mach} H.~Mach {\it et al.}, Phys. Rev. C {\bf 41}, R2469 (1990).
\bibitem{Pitz} H.H.~Pitz {\it et al.}, Nucl. Phys. A {\bf 509}, 587 (1990).
\bibitem{Wollersheim} H.J.~Wollersheim {\it et al.},
Nucl. Phys. A {\bf 556}, 261 (1993).
\bibitem{Raman} S.~Raman {\it et al.},
At. Data Nucl. Data Tables {\bf 36}, 1 (1987).
\bibitem{Cline} D.~Cline, Nucl. Phys. A {\bf 557}, 615c (1993).
\bibitem{Landau} L.D.~Landau and E.M.~Lifschitz,  {\it Quantenmechanik}
(Akademie--Verlag, Berlin, 1965) p.185.
\end{thebibliography}
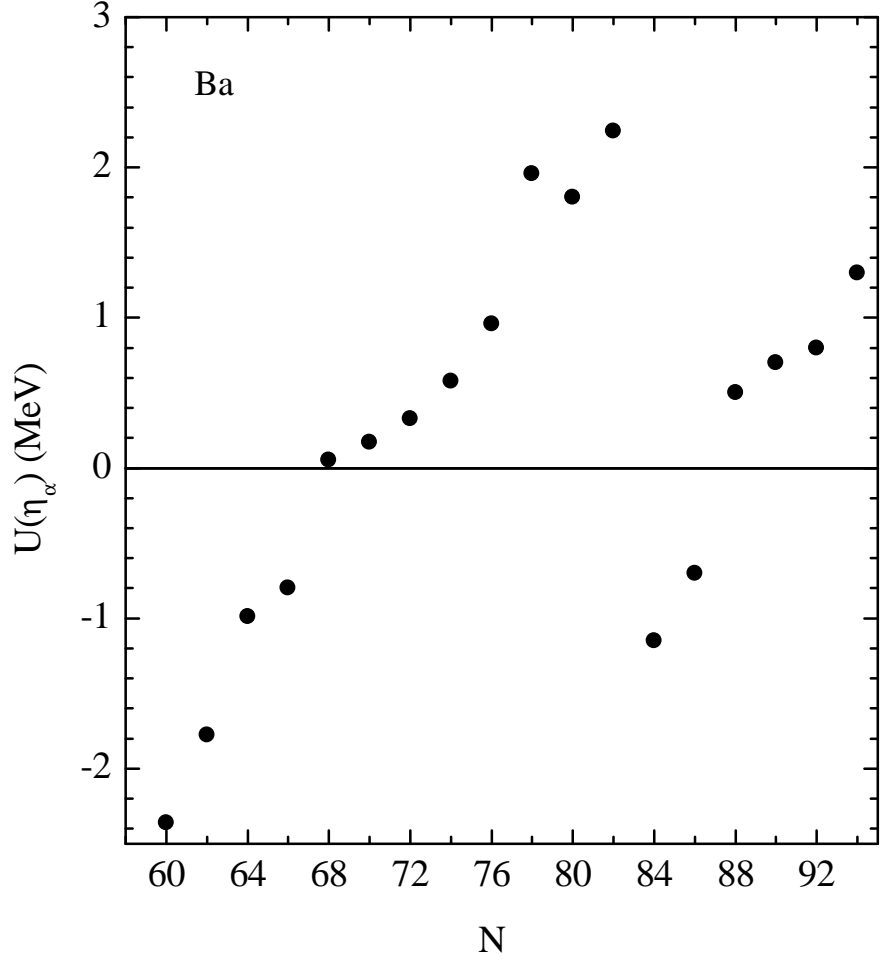
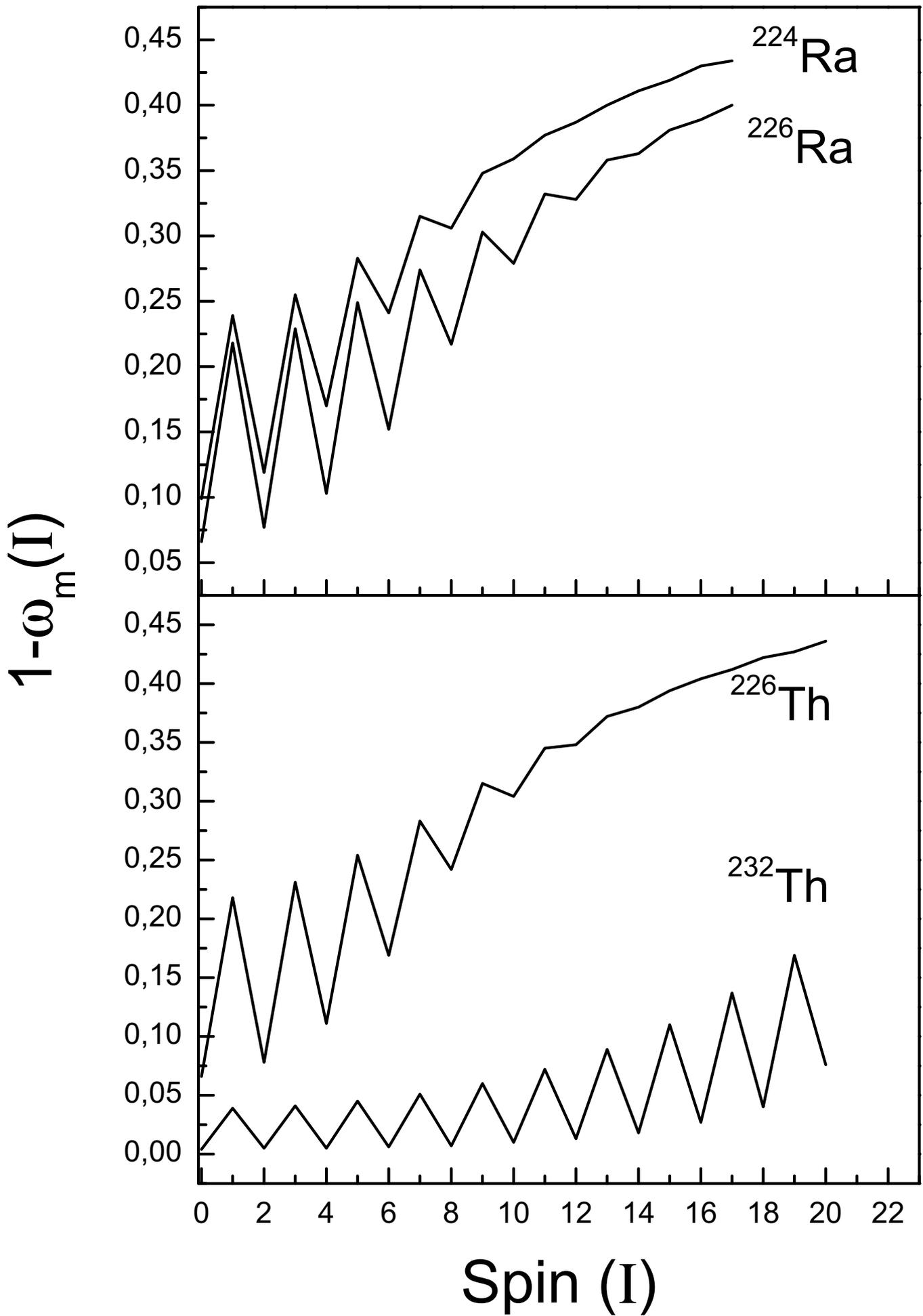
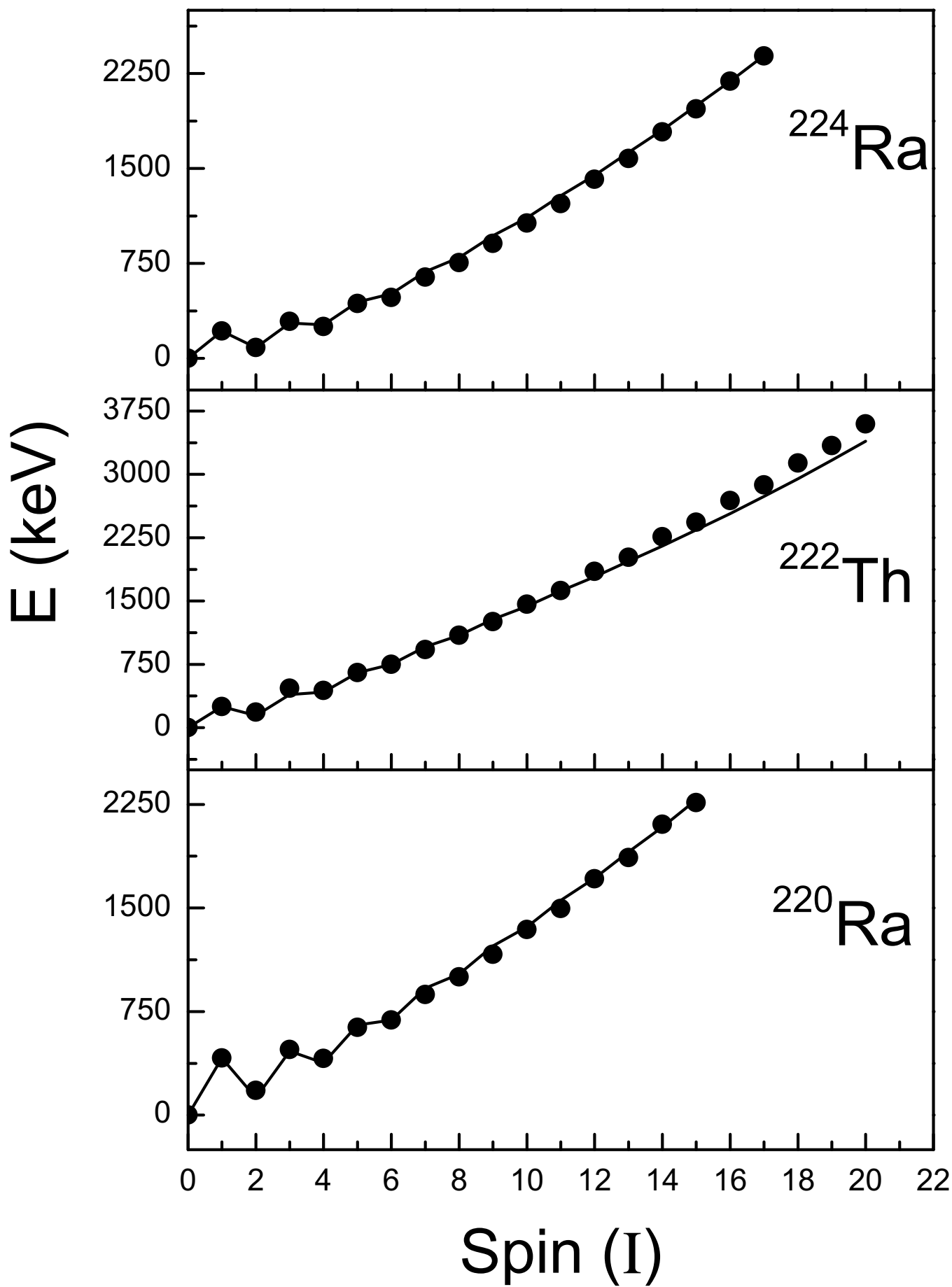
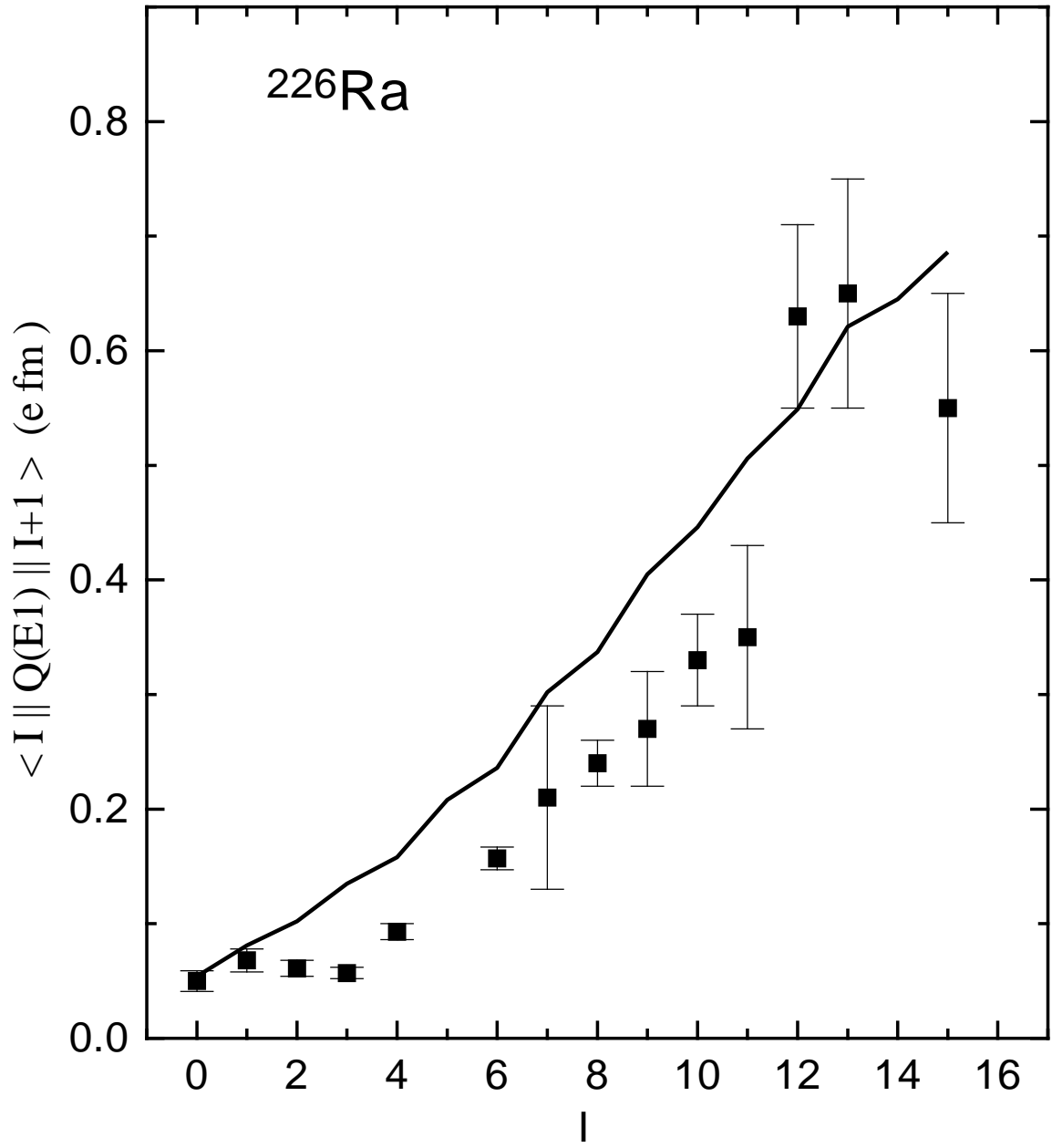
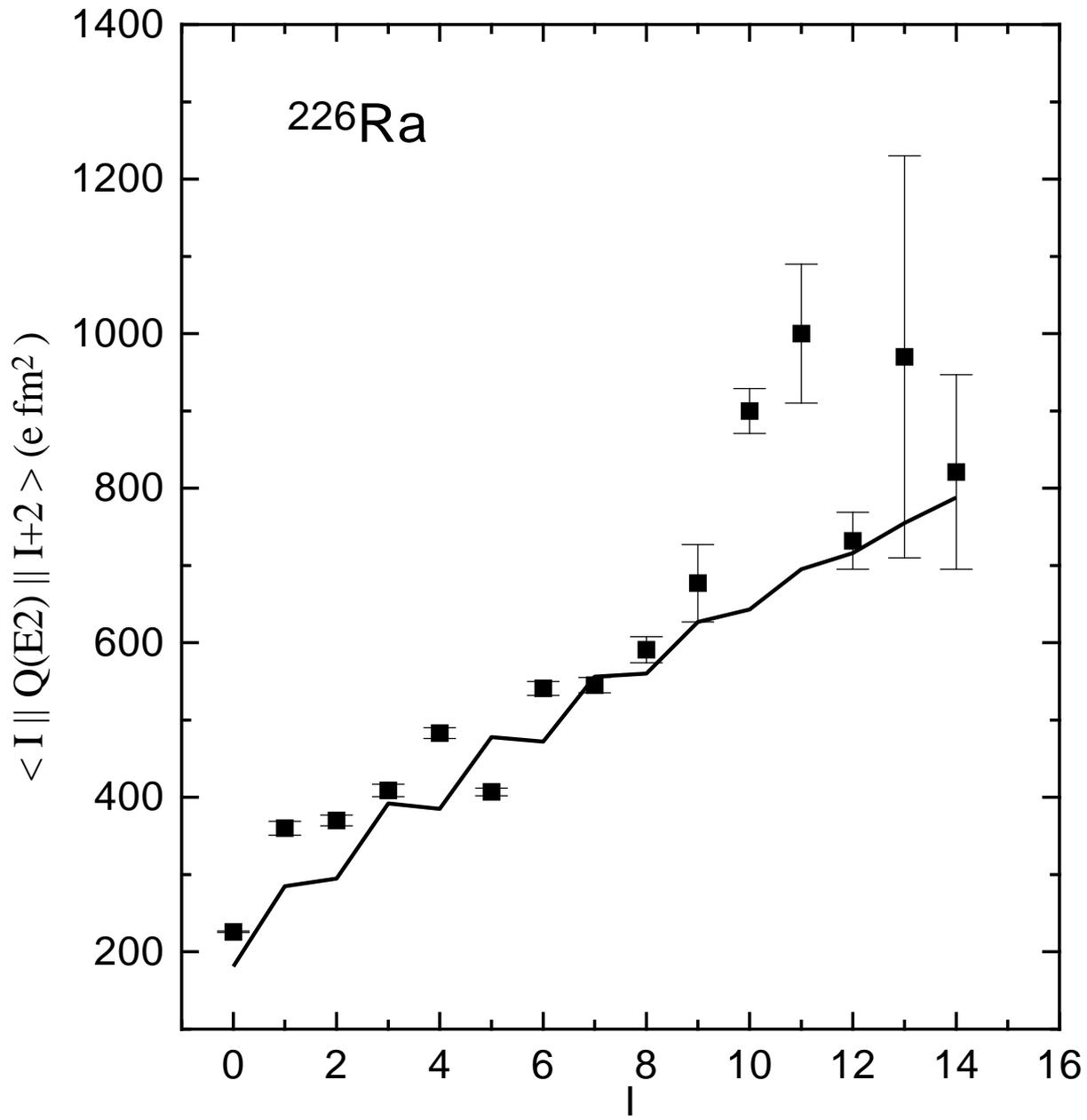
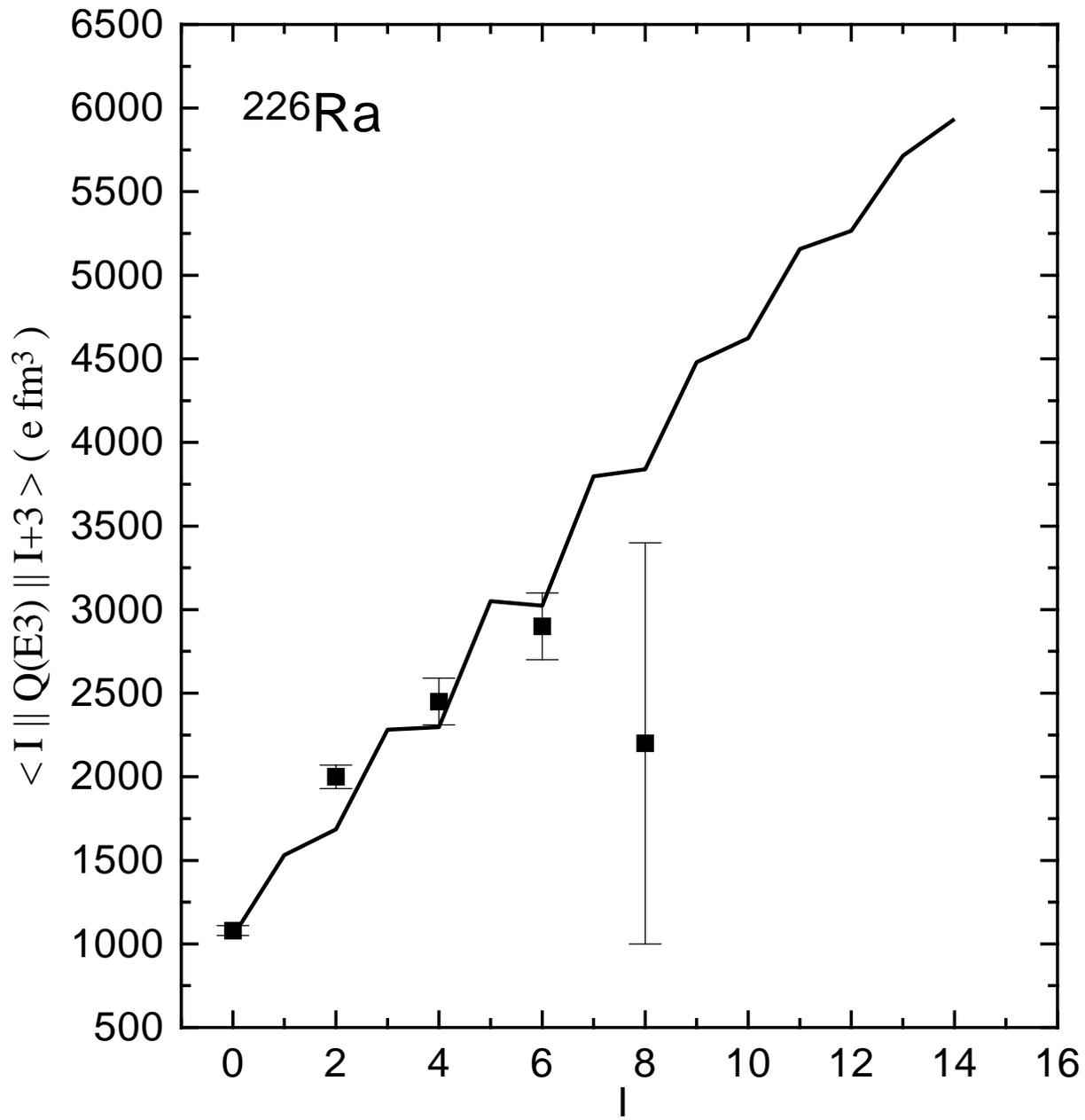
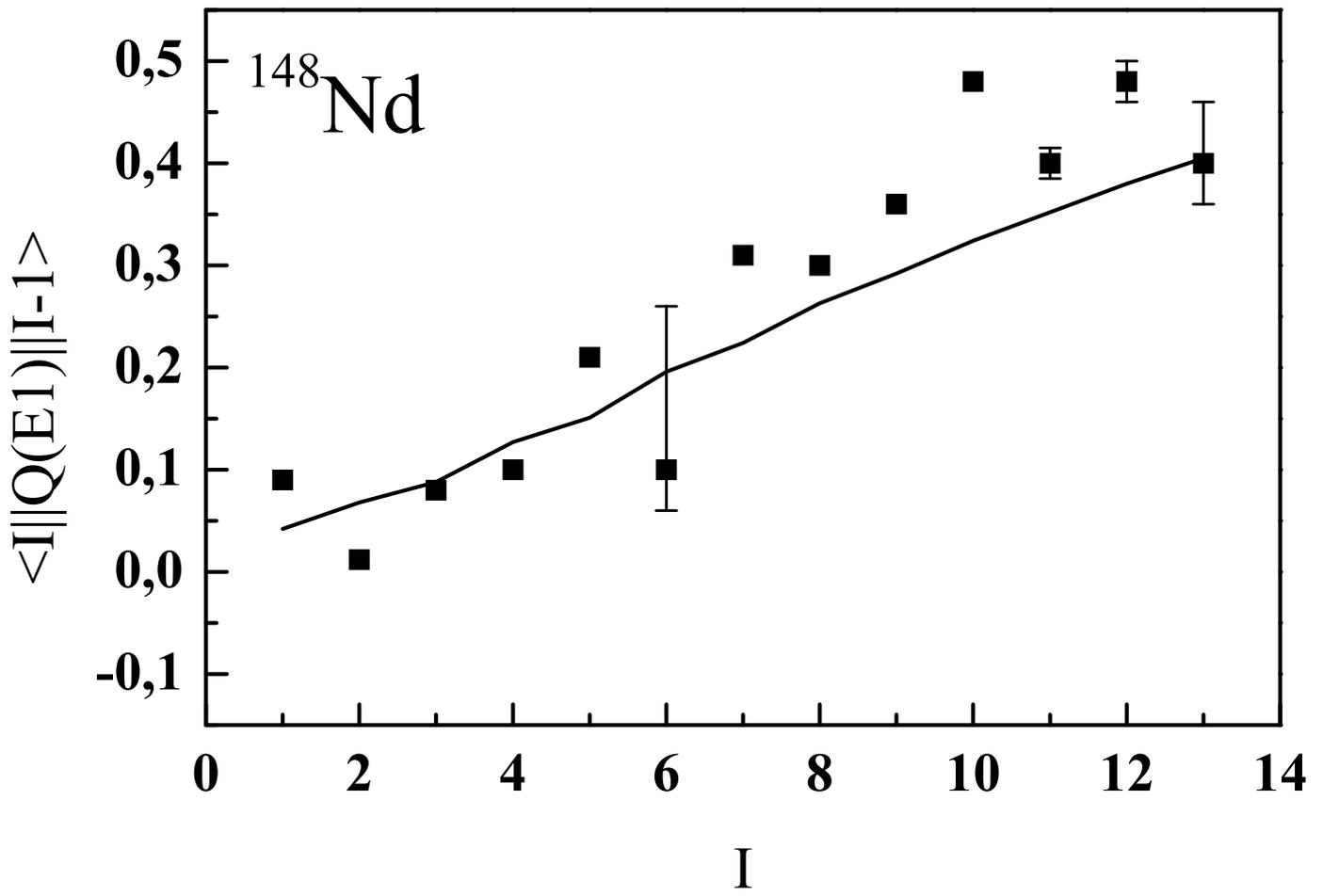
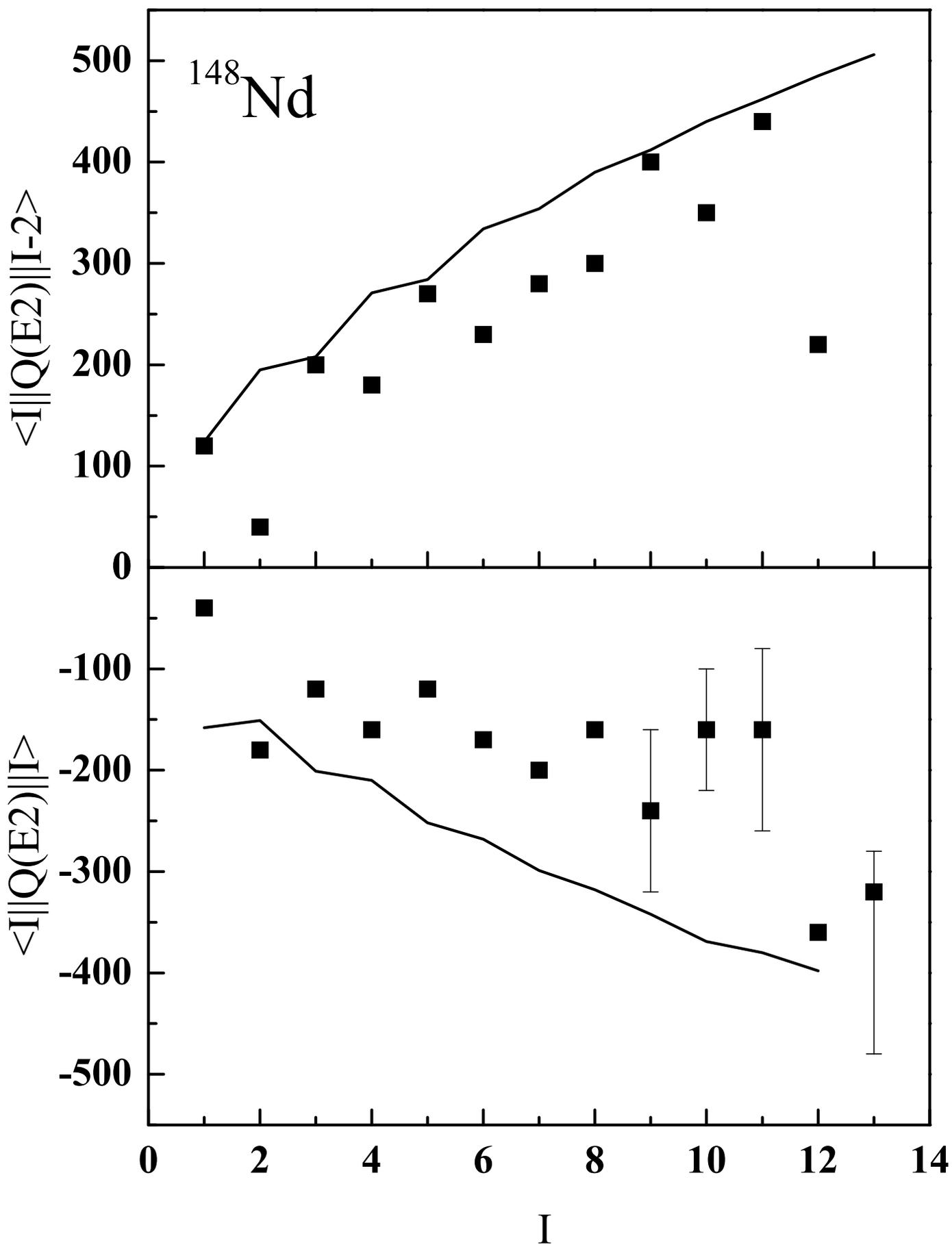
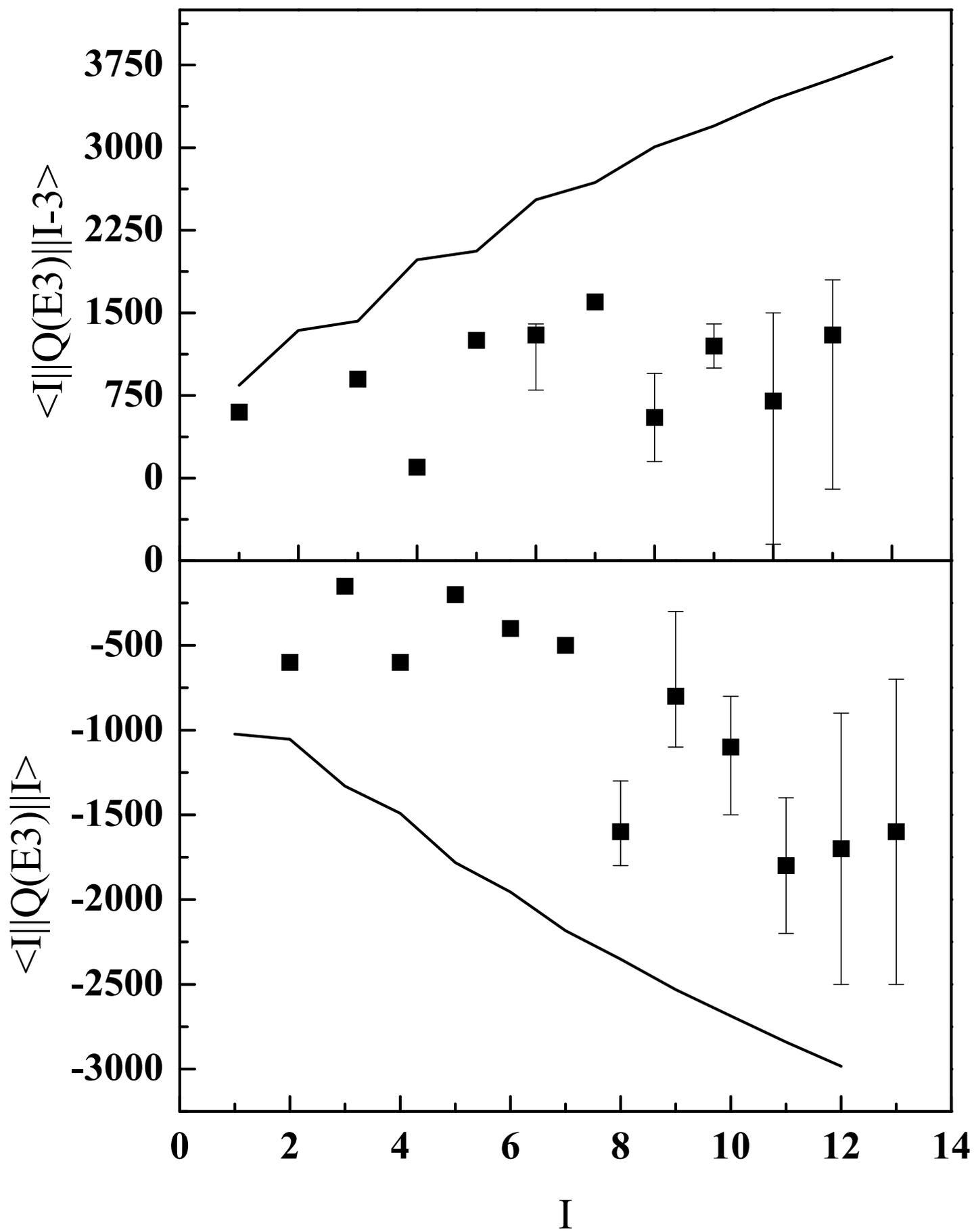
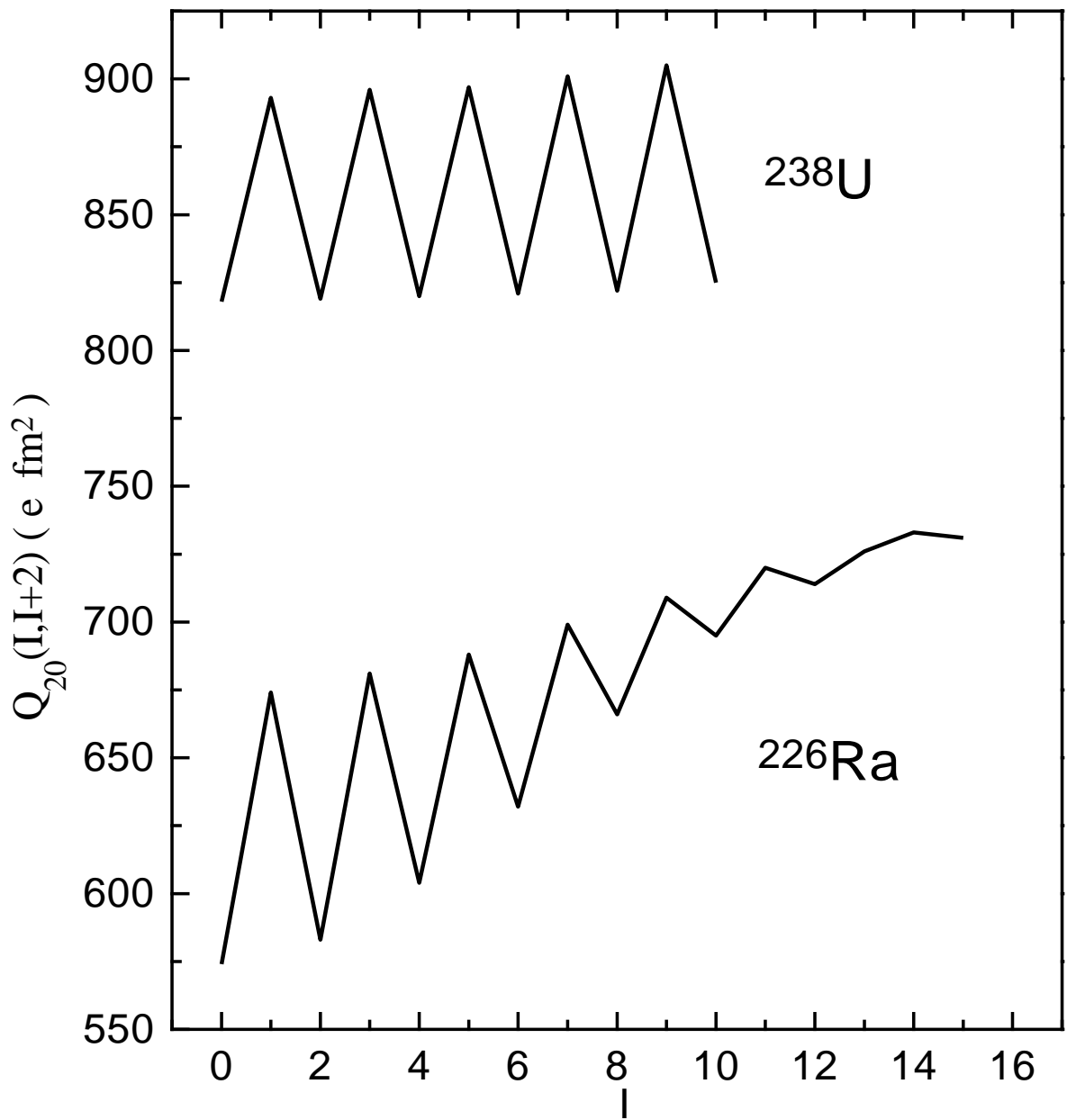
\end{document}